\input{aipcheck}

\documentclass[
    ,final            
  ]
  {aipproc}

\layoutstyle{8x11single}

\begin{document}

\title{Nonlinear dynamics of recollisions in the double ionization processes of atoms in strong fields}


\author{F.~Mauger}{
  address={Centre de Physique Th\'eorique, CNRS -- Aix-Marseille Universit\'e, Luminy - case 907, F-13288 Marseille cedex 09, France }
}

\author{C.~Chandre}{
  address={Centre de Physique Th\'eorique, CNRS -- Aix-Marseille Universit\'e, Luminy - case 907, F-13288 Marseille cedex 09, France }
}

\author{T.~Uzer}{
  address={School of Physics, Georgia Institute of Technology, Atlanta, GA 30332-0430, USA}
}

\begin{abstract}
    Double ionization processes triggered by intense linearly polarized laser fields have revealed the paramount role of electron-electron collisions. It has been shown~\cite{Maug10} that the correlated dynamics of such electronic collisions can be modeled by a two dimensional symplectic map which captures the bare essentials of the energy exchange processes occurring during a recollision between the two electrons. We investigate linear and nonlinear properties of this map and connect them to the dynamical organization of phase space and related statistical data.
\end{abstract}

\maketitle




\section{Introduction}
Atoms and molecules subjected to intense linearly polarized laser fields have demonstrated the dramatic impact of electron-electron correlation in the double ionization process. The ionization channel by which the two electrons collide before ionizing (called nonsequential double ionization) has been shown to be the dominant channel for a significant range of laser intensity; indeed the corresponding double ionization yields were found to be several orders of magnitude higher than the sequential mechanism predicts~\cite{Fitt92,Kond93,Walk94,Laro98,Webe00,Corn00,Guo01,DeWi01,Ruda04}. The recollision~\cite{Cork93,Scha93} in which an electron is first ionized and then hurled back to the core upon sign reversal of the laser field and collides with the other electron, has been identified as the ``keystone of strong field physics''~\cite{Beck08}. It has been confronted successfully with experiments~\cite{Fitt92,Kond93,Walk94,Laro98,Webe00,Corn00,Guo01,DeWi01,Ruda04,Ivan05}, quantum~\cite{Lein00, Wats97}, semi-classical simulations~\cite{Fu01, Chen03, Brab96}. Surprisingly, even classical simulations~\cite{Panf03, Fu01, Sach01, Ho05_1, Ho05_2, Panf02, Liu07, Maug09, Maug10} manage to reproduce such results and identify a classical recollision scenario as the dominant double ionization channel in the relevant intensity range. The reason for this success is the presence of sufficient electron-electron correlations in all such models. 

Within the dipole approximation and using soft Coulomb potentials~\cite{Java88, Ho05_1, Ho05_2, Panf02}, the Hamiltonian of a one dimensional helium atom driven by an intense laser field is written in atomic units (a.u.) as
\begin{equation} \label{eq:Hamiltonian}
   H \left( x, y, p_{x}, p_{y}, t \right) = \frac{p_{x}^{2}}{2} + \frac{p_{y}^{2}}{2} + \frac{1}{\sqrt{ \left( x-y \right)^{2} + 1}} -\frac{2}{\sqrt{x^{2}+1}} - \frac{2}{\sqrt{y^{2}+1}} + (x+y) E_{0} \sin  \omega t,
\end{equation}
where $x$ and $y$ are respectively the position of the two electrons (along the axis of polarization) and $p_{x}$ and $p_{y}$ their (canonically) conjugate momenta. The laser parameters are the amplitude of the field $E_{0}$, and its frequency $\omega$. 

Despite the two electrons are treated equally in Hamiltonian~(\ref{eq:Hamiltonian}), a careful analysis of typical trajectories reveals dramatically different behaviors for each of them~\cite{Maug09}. One electron (the outer electron) is quickly picked up by the field and ionized while the other one (the inner electron) is left within a competition between the laser excitation and the (soft) Coulomb attraction from the core. This asymmetry is explained by the dynamical organization of phase space by a couple of periodic orbits in the absence of the field~\cite{Maug09}. From this competition between the laser and the Coulomb attraction, mainly two outcomes are possible
\begin{itemize}
   \item If the inner electron is sufficiently far away from the nucleus (in a region referred as the unbound area), the laser excitation imposes the dynamics and the electron ionizes quickly.
   \item If the inner electron is sufficiently close to the nucleus (in a region referred as the bound area), it is trapped on an invariant torus and has no chance to ionize by itself.
\end{itemize}
The respective size of these bound and unbound regions depend on the parameters of the laser field. 
To ionize an electron in the bound area one has to find a sufficient amount of energy to move to the unbound area. This can be achieved when the outer electron comes back to the core region and collides, i.e.\ exchanges energy, with the inner electron. Due to the periodicity of the laser field, the outer electron is brought back to the core nearly periodically since the laser is linearly polarized. Therefore the recollisions can be modeled as kicks experienced periodically by the inner electron. 

\section{Reduced model for the recollision dynamics}

Since the dynamics in the bound area is filled by invariant tori, it is natural to perform a change of coordinates into action--angle variables. This (canonical) change of coordinates associates a constant action with each torus on which the inner electron is temporarily confined. Thus the jumps from one torus to another, experienced by the inner electron during the recollision process, is reduced to a change of action in the new set of coordinates. By statistical analysis of the recollision process together with theoretical reduced models for the inner and outer electrons, a reduced kicked Hamiltonian for the dynamics of the inner electron driven by recollisions was proposed~\cite{Maug10}. Its integration yields to the following symplectic map~:
\begin{equation} \label{eq:Mapping}
   \begin{array}{lcl}
      A_{n+1}       & = & A_{n}/(1- \varepsilon \sin \varphi_{n}), \\
      \varphi_{n+1} & = & \varphi_{n} + T\sqrt{2} \exp(a A_{n+1})+\varepsilon\cos \varphi_n,
   \end{array}
\end{equation}
where $A_{n}$ and $\varphi_{n}$ are respectively the action and angle variables of the inner electron right before the $n^{\rm th}$ kick. The parameter $T$ is half of the period of the laser field, i.e.\ $T=\pi/\omega$. The parameter $\varepsilon$, which only depends on the parameters of the laser field (here its intensity since we have kept the frequency fixed), simulates the variation of recollision efficiency as the laser intensity is varied and it is given by
$$
   \varepsilon = \frac{\alpha E_{0}^{2}}{1 + \beta E_{0}^{3}},
$$
where the parameter $\alpha= \kappa/\left(4\sqrt{2}\omega^{2}\right)$ with $\kappa=3.17\ldots$ can be computed analytically and the parameter $\beta=3\kappa^2\gamma'^{2}/(16\eta L^4\omega^2)$. Since there is an adjustable parameter corresponding to the interaction length between the two electrons ($L$), $\beta$ is computed using a numerical fit (for more details, see Ref.~\cite{Maug10}). The intensity of the laser is proportional to the square of the amplitude and we note that the efficiency scales proportionally to the intensity for low intensities, and it decreases as the inverse of its square root for larger intensities. The parameter $a = -9\sqrt{2}/16$ is obtained using a series expansion of the dynamics around the nucleus.

Because of the finite size of the bound area, there exists a boundary invariant torus (defined as the outermost invariant torus analytically conjugated to a rotation $(A(t),\varphi(t))=(A_{\rm m},\Omega t+\varphi_{\rm m})$). It defines a maximum action $A_{\rm m}$ for the validity of the reduced model~(\ref{eq:Mapping}). Since the size of the bound area depends on the intensity, so does $A_{\rm m}$. It enables one to predict a ionization probability of the inner electron driven by recollisions~: An electron whose dynamics is modeled by the map~(\ref{eq:Mapping}) is considered as ionized a soon as its action $A_n$ becomes larger than $A_{\rm m}$ after the $n^{\rm th}$ kick where the iteration of the map~(\ref{eq:Mapping}) is stopped.

Compared to statistical results on the double ionization probabilities obtained with the full Hamiltonian~(\ref{eq:Hamiltonian}), the reduced kicked dynamics described by the map~(\ref{eq:Mapping}) agrees very well in the whole range of intensities where nonsequential double ionization is the dominant channel. This reduction simplifies drastically the complexity of the analysis~: One ends up with an explicit discrete map (with two variables) instead of a continuous flow with two and a half degrees of freedom. In what follows we study some properties of the map~(\ref{eq:Mapping}) to connect them with the dynamics of the inner electron subjected to periodic collisions with the outer one.

\section{Analysis of the map~(\ref{eq:Mapping})}

\subsection{Asymptotic behavior}

Various kicked models have been developed for atomic systems such as Rydberg atoms subjected to microwave fields~\cite{Casa88, ChaosAtomPhys}. With cold atoms such kicked maps have enabled one to identify and understand accelerator modes~\cite{Buch06}. For the problem at hand, because of the term $T\sqrt{2} \exp(a A_{n+1})$ in Eq.~(\ref{eq:Mapping}), the acceleration condition is asymptotic, when $A_{n} \rightarrow \infty$ (since $a<0$). This asymptotic limit gives two possible values of the angle for the accelerator mode condition $\varphi_{n} = \pm \pi/2$ such that $\varphi_{n+1} \approx \varphi_{n}$ and $A_{n+1} = A_{n}/\left(1 \mp \varepsilon\right)$ in Eq.~(\ref{eq:Mapping}). Here $\varepsilon$ is a positive constant and depending on the sign in the previous equation, $A_{n+1}$ is either larger or smaller than $A_{n}$. Since the condition holds when $A_{n} \rightarrow \infty$, only $\varphi_{n} = \pi/2$ corresponds to an asymptotic acceleration.

This acceleration is clearly visible on a (complete) phase space of the mapping~(\ref{eq:Mapping}) and is associated with a burst toward very large actions around the angle $\pi/2$. However, this condition is asymptotic and thus holds only for $\left|a\right| A_{n} \gg 1$ or equivalently $A_{n} \gg 16/\left(9 \sqrt{2}\right) \approx 1.25$ which exceeds the critical action $A_{\rm m}$ for ionization of the inner electron all over the range of considered intensities. As a consequence, accelerator modes do not contribute significantly for the problem at hand. They would contribute for low intensities and for rather long laser pulses. 

\subsection{Linear stability}

\begin{figure}
  \includegraphics[width=.29\linewidth]{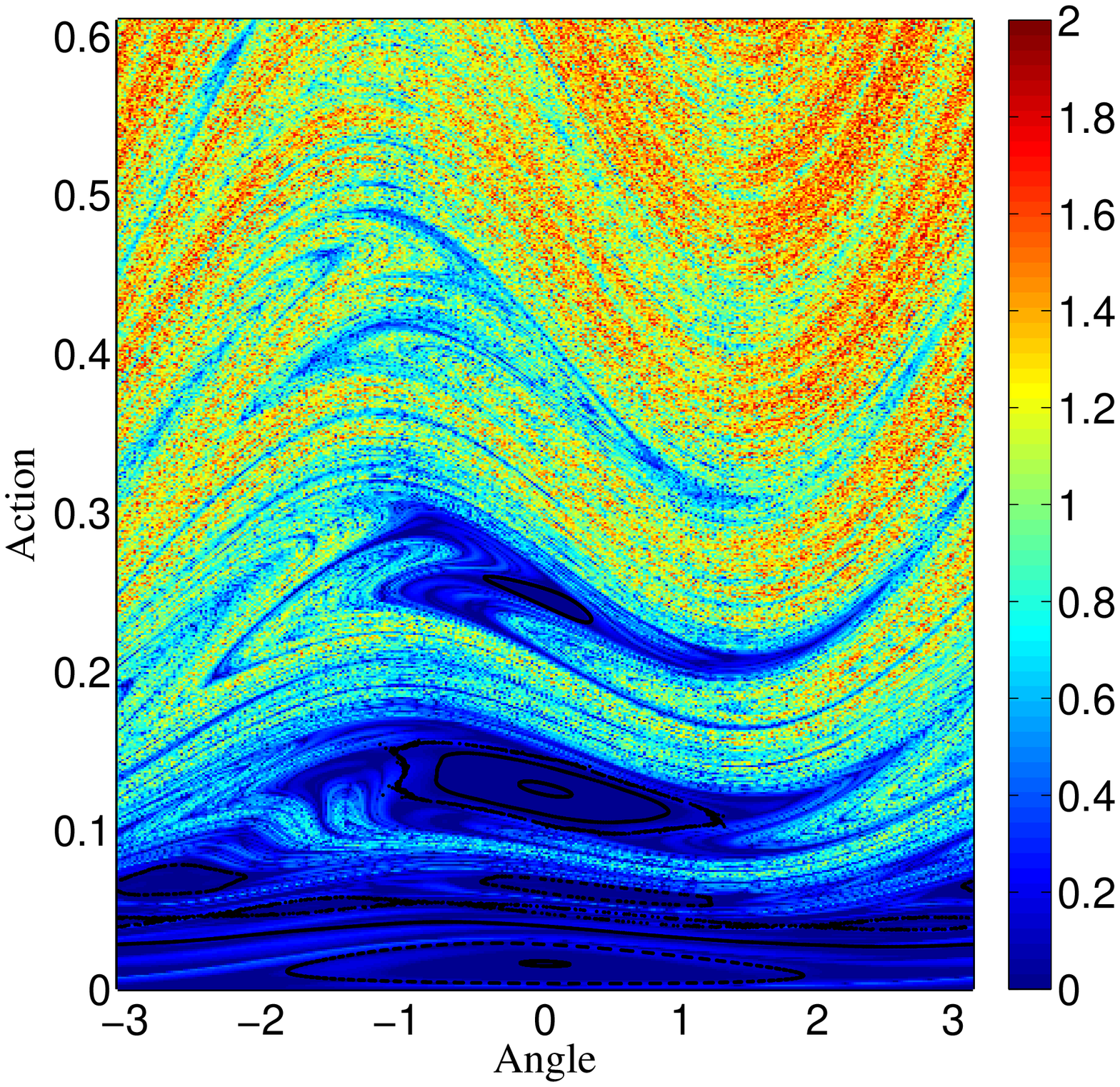}
  \includegraphics[width=.29\linewidth]{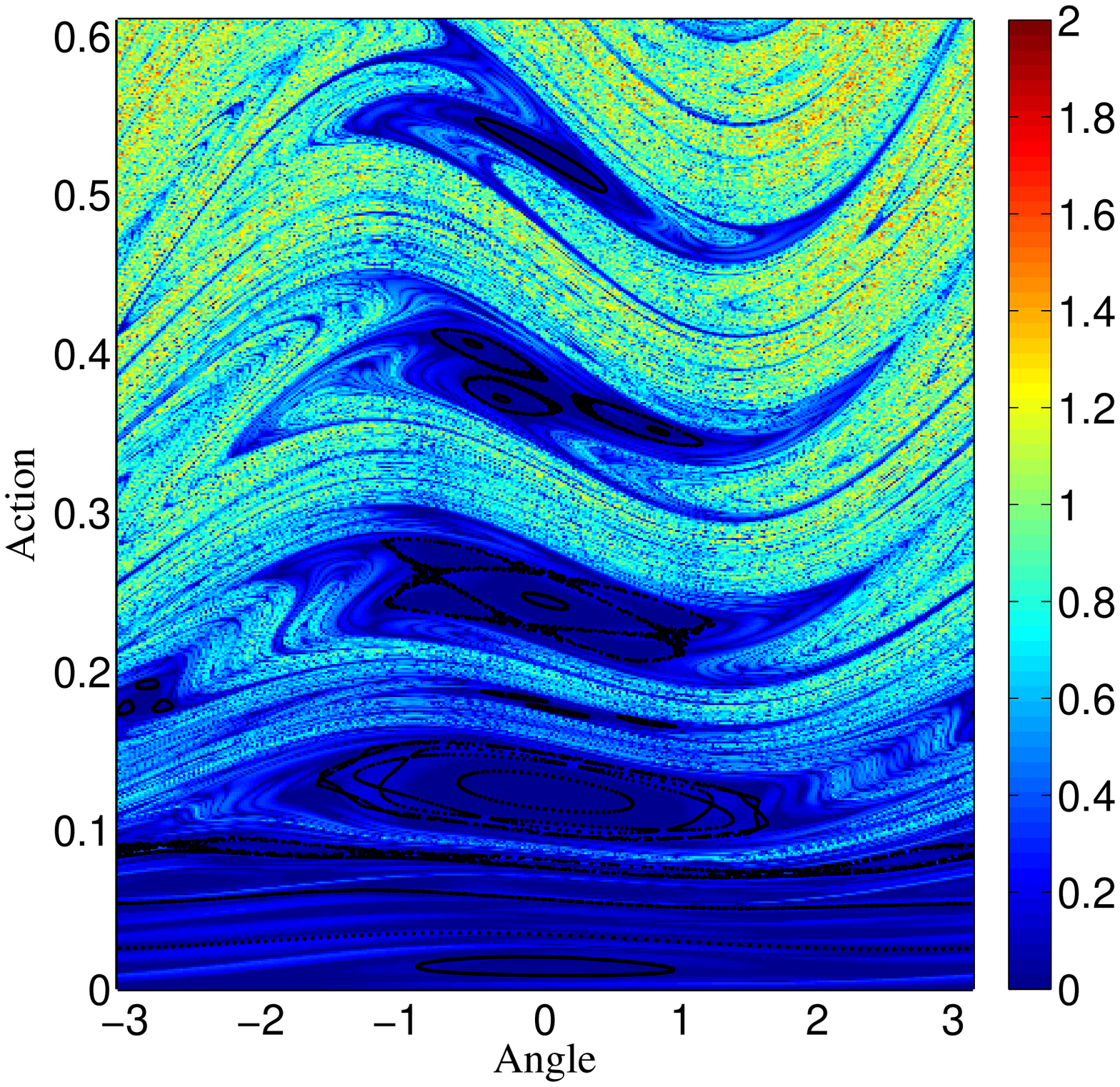}
  \includegraphics[width=.29\linewidth]{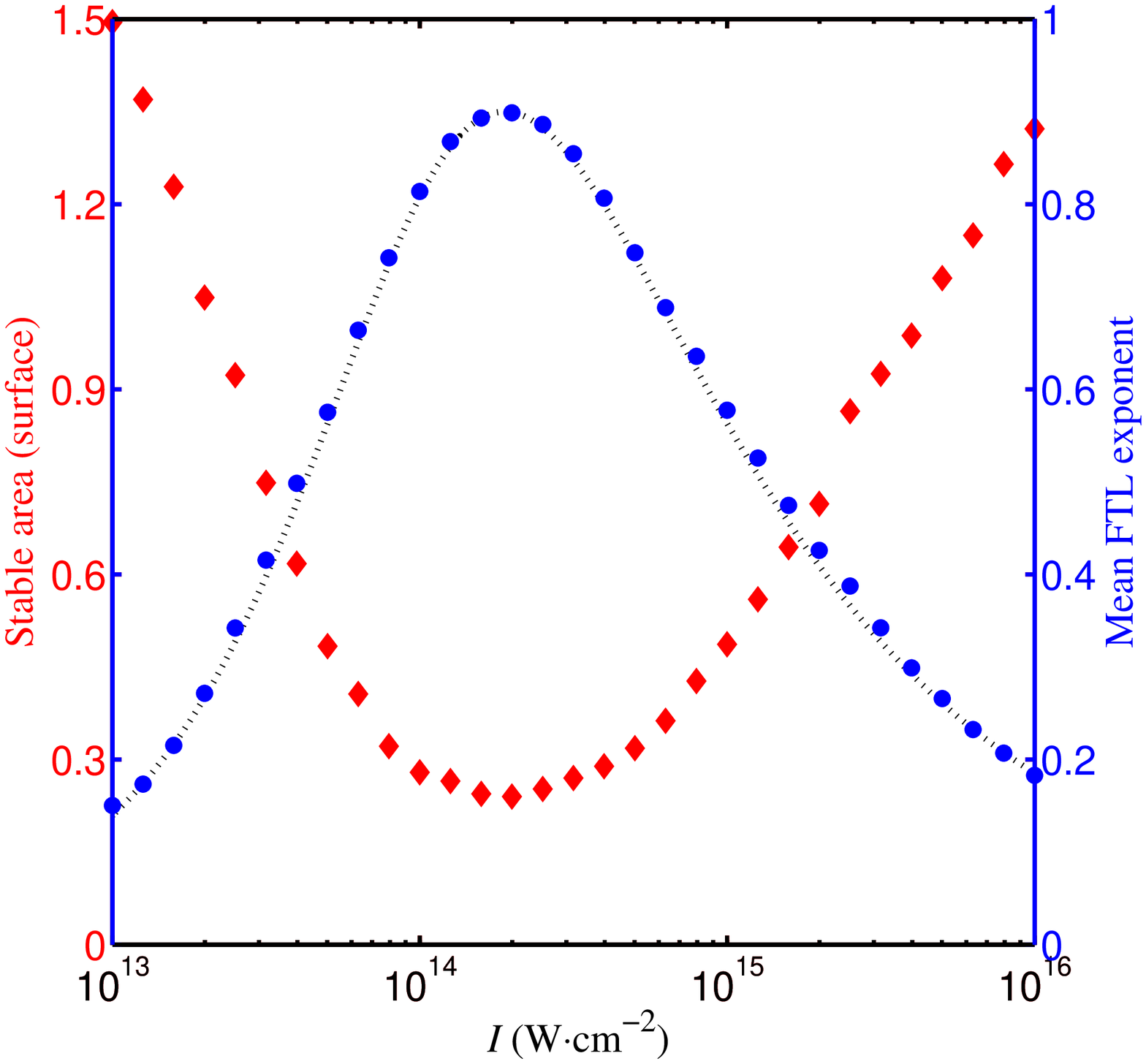}
  \caption{\label{fig:LyapunovMap}
  Two left panels: Finite Time Lyapunov maps of the dynamics given by Eq.~(\ref{eq:Mapping}) for laser intensities $3 \times 10^{14} \ {\rm W}\cdot{\rm cm}^{-2}$ (left panel) and $10^{15} \ {\rm W}\cdot{\rm cm}^{-2}$ (middle panel). For comparison, we also display some trajectories in the stable regions for a large number of iterations of the map~(\ref{eq:Mapping}). Right panel: Linear stability indicators as given by FTL exponents as a function of the laser field intensity. We compute the surface of stable phase space (red diamonds, left hand horizontal axis) and the mean FTL exponent after 10 iterations of the map~(\ref{eq:Mapping}) (blue circles, right hand horizontal axis). For comparison, we also display the variation of $\varepsilon$ with intensity (dotted black curve, arbitrary units). Stable initial conditions are those for which the associate FTL exponent is small at the end of the iteration of the map.}
\end{figure}

A quick inspection of trajectories of the map~(\ref{eq:Mapping}) shows regular and chaotic areas in phase space. As the efficiency of the kicks (e.g., $\varepsilon$) increases, such phase-space portraits become more sparse: elliptic (stable) islands shrink and the diffusion in the chaotic layer increases~\cite{Maug10}. Using linear stability analysis such as given by finite time Lyapunov (FTL) exponents~\cite{chaosbook} we analyze the dynamics in the accessible phase-space for the inner electron. The way to do this analysis is through FTL maps which indicate the value of the exponent as a function of the initial condition in phase-space. Two FTL maps are depicted in Fig.~\ref{fig:LyapunovMap} and compared with trajectories in the elliptic islands. FTL maps clearly highlight the coexistence of stable (small FTL exponent, blue regions) and chaotic (large FTL exponent, red regions) areas. In the chaotic layer, we also note the decrease of the FTL exponent when $\varepsilon$ decreases (by comparing color maps between the figures for the two laser intensities $3 \times 10^{14} \ {\rm W}\cdot{\rm cm}^{-2}$ and $10^{15} \ {\rm W}\cdot{\rm cm}^{-2}$)

If the inner electron has initial conditions in a stable region (e.g., in one of the elliptic islands), then it will be impossible for the outer electron to dislodge the inner one regardless of the number of kicks (which is related to the pulse duration). These regions in phase space do not lead to double ionization. 
We investigate further the correlation between the (linear) stability of the map~(\ref{eq:Mapping}) and the efficiency $\varepsilon$ of the kicks, namely we compute the phase space surface for which the inner electron is linearly stable (small FTL exponent) and the mean value of the FTL exponent in the chaotic layer (large FTL exponent). All over the range of intensities where the mapping model can be used to simulate the collision dynamics, we see that the variation of the area of stable motion is correlated to the efficiency of the kicks~: This area increases when the efficiency of the kicks decreases which is expected. We also note the perfect matching (up to a multiplicative constant) between the mean value of the FTL exponent ($m_{FTI}$) in the unstable region and the efficiency of the kicks (i.e.\ $\varepsilon \propto m_{FTI}$). As the FTL exponents are indicators of the (un)stability of trajectories, it confirms what we have already qualitatively observed looking at phase-space portraits~\cite{Maug10}.

\section{Conclusion}

We have related linear and nonlinear properties of a two-dimensional symplectic map with the electron--electron collision dynamics driven by intense and short laser pulses. Accelerator modes, only accessible in the asymptotic limit of large actions are shown not to play a role in the dynamical process. The linear stability analysis shows the strong correlation between the regularity of the dynamics and the efficiency of the kicks~: The stable area shrinks with an increase of the efficiency of the kicks, and the mean Lyapunov exponent in the unstable region is proportional to the efficiency of the kicks.


\begin{theacknowledgments}
  We acknowledge useful discussions with P.~J. Morrison and A. Kamor. CC and FM acknowledge financial support from the PICS program of the CNRS. This work is partially funded by NSF.
\end{theacknowledgments}



\bibliographystyle{aipproc}   


\begin{thebibliography}{40}
\expandafter\ifx\csname natexlab\endcsname\relax\def\natexlab#1{#1}\fi
\providecommand{\enquote}[1]{``#1''}
\expandafter\ifx\csname url\endcsname\relax
  \def\url#1{\texttt{#1}}\fi
\expandafter\ifx\csname urlprefix\endcsname\relax\def\urlprefix{URL }\fi
\providecommand{\eprint}[2][]{\url{#2}}


\bibitem[Mauger et~al.(2010{\natexlab{a}})]{Maug10}
F.~Mauger, C.~Chandre, and T.~Uzer, \emph{Phys.~Rev.~Lett.} \textbf{104},
  043005 (2010{\natexlab{a}}); \emph{Phys.~Rev.~A} \textbf{81}, 063425
  (2010{\natexlab{b}}).
  
\bibitem[Becker and Rottke(2008)]{Beck08}
W.~Becker, and H.~Rottke, \emph{Contemporary Physics} \textbf{49}, 199--223
  (2008).

\bibitem[Fittinghoff et~al.(1992)]{Fitt92}
D.~N. Fittinghoff, {\it et al.}\ 
  \emph{Phys.~Rev.~Lett.} \textbf{69}, 2642--2645 (1992).

\bibitem[Kondo et~al.(1993)]{Kond93}
K.~Kondo, {\it et al.}\ 
  \emph{Phys.~Rev.~A} \textbf{48}, R2531--R2533 (1993).

\bibitem[Walker et~al.(1994)]{Walk94}
B.~Walker, {\it et al.}\ 
  \emph{Phys.~Rev.~Lett.} \textbf{73}, 1227--1230 (1994).

\bibitem[Larochelle et~al.(1998)]{Laro98}
S.~Larochelle, A.~Talebpour, and S.~L. Chin, \emph{J.~Phys.~B.} \textbf{31},
  1201--1214 (1998).

\bibitem[Weber et~al.(2000)]{Webe00}
T.~Weber, {\it et al.}\ 
  \emph{Nature}
  \textbf{405}, 658--661 (2000).

\bibitem[Cornaggia and Hering(2000)]{Corn00}
C.~Cornaggia, and P.~Hering, \emph{Phys.~Rev.~A} \textbf{62}, 023403 (2000).

\bibitem[Guo and Gibson(2001)]{Guo01}
C.~Guo, and G.~N. Gibson, \emph{Phys.~Rev.~A} \textbf{63}, 040701 (2001).

\bibitem[DeWitt et~al.(2001)]{DeWi01}
M.~J. DeWitt, E.~Wells, and R.~R. Jones, \emph{Phys.~Rev.~Lett.} \textbf{87},
  153001 (2001).

\bibitem[Rudati et~al.(2004)]{Ruda04}
J.~Rudati, {\it et al.}\ 
  \emph{Phys.~Rev.~Lett.} \textbf{92}, 203001 (2004).

\bibitem[Corkum(1993)]{Cork93}
P.~B. Corkum, \emph{Phys. Rev. Lett.} \textbf{71}, 1994--1997 (1993).

\bibitem[Schafer et~al.(1993)]{Scha93}
K.~J. Schafer, {\it et al.}\ 
  \emph{Phys. Rev.  Lett.} \textbf{70}, 1599--1602 (1993).

\bibitem[Becker and Faisal(1996)]{Beck96}
A.~Becker, and F.~H.~M. Faisal, \emph{J.~Phys.~B.} \textbf{29} (1996).

\bibitem[Kopold et~al.(2000)]{Kopo00}
R.~Kopold, {\it et al.}\ 
  \textbf{85}, 3781--3784 (2000).

\bibitem[Lein et~al.(2000)]{Lein00}
M.~Lein, E.~K.~U. Gross, and V.~Engel, \emph{Phys.~Rev.~Lett.} \textbf{85},
  4707--4710 (2000).

\bibitem[Sacha and Eckhardt(2001)]{Sach01}
K.~Sacha, and B.~Eckhardt, \emph{Phys. Rev. A} \textbf{63}, 043414 (2001).

\bibitem[Fu et~al.(2001)]{Fu01}
L.-B. Fu, {\it et al.}\ 
  043416 (2001).

\bibitem[Panfili and Eberly(2001)]{Panf01}
R.~Panfili, and J.~H. Eberly, \emph{Opt.~Express} \textbf{8}, 431--435 (2001).

\bibitem[Barna and Rost(2003)]{Barn03}
I.~Barna, and J.~Rost, \emph{Eur.~Phys.~J.~D} \textbf{27}, 287 (2003).

\bibitem[Colgan et~al.(2004)]{Colg04}
J.~Colgan, M.~S. Pindzola, and F.~Robicheaux, \emph{Phys.~Rev.~Lett.}
  \textbf{93}, 053201 (2004).

\bibitem[Ho et~al.(2005)]{Ho05_1}
P.~J. Ho, {\it et al.}\ 
  \textbf{94}, 093002 (2005).

\bibitem[Ho and Eberly(2005)]{Ho05_2}
P.~J. Ho, and J.~H. Eberly, \emph{Phys. Rev. Lett.} \textbf{95}, 193002 (2005).

\bibitem[Ruiz et~al.(2005)]{Ruiz05}
C.~Ruiz, L.~Plaja, and L.~Roso, \emph{Phys.~Rev.~Lett.} \textbf{94}, 063002
  (2005).

\bibitem[Horner et~al.(2007)]{Horn07}
D.~A. Horner, {\it et al.}\ 
  \emph{Phys.~Rev.~A} \textbf{76}, 030701 (2007).

\bibitem[Prauzner-Bechcicki et~al.(2007)]{Prau07}
J.~S. Prauzner-Bechcicki, {\it et al.}\ 
  \emph{Phys.~Rev.~Lett.} \textbf{98}, 203002 (2007).

\bibitem[Feist et~al.(2008)]{Feis08}
J.~Feist, {\it et al.}\ 
  and J.~Burgd\"orfer, \emph{Phys.~Rev.~A} \textbf{77}, 043420 (2008).

\bibitem[Ivanov et~al.(2005)]{Ivan05}
M.~Y. Ivanov, M.~Spanner, and O.~Smirnova, \emph{J.~Mod.~Opt.} \textbf{52},
  165--184 (2005).

\bibitem[Watson et~al.(1997)]{Wats97}
J.~B. Watson, {\it et al.}\ 
  \emph{Phys.~Rev.~Lett.} \textbf{78}, 1884--1887 (1997).

\bibitem[Chen et~al.(2003)]{Chen03}
J.~Chen, J.~H. Kim, and C.~H. Nam, \emph{J.~Phys.~B.} \textbf{36}, 691 (2003).

\bibitem[Brabec et~al.(1996)]{Brab96}
T.~Brabec, M.~Y. Ivanov, and P.~B. Corkum, \emph{Phys.~Rev.~A} \textbf{54},
  R2551--R2554 (1996).

\bibitem[Panfili and Liu(2003)]{Panf03}
R.~Panfili, and W.-C. Liu, \emph{Phys. Rev. A} \textbf{67}, 043402 (2003).

\bibitem[Panfili et~al.(2002)]{Panf02}
R.~Panfili, S.~L. Haan, and J.~H. Eberly, \emph{Phys. Rev. Lett.} \textbf{89},
  113001 (2002).

\bibitem[Liu et~al.(2007)]{Liu07}
J.~Liu, D.~F. Ye, J.~Chen, and X.~Liu, \emph{Phys.~Rev.~Lett.} \textbf{99},
  013003 (2007).


\bibitem[Mauger et~al.(2009{\natexlab{a}})]{Maug09}
F.~Mauger, C.~Chandre, and T.~Uzer, \emph{Phys.~Rev.~Lett.} \textbf{102},
  173002 (2009{\natexlab{a}}); \emph{J.~Phys.~B.} \textbf{42}
  (2009{\natexlab{b}}).

\bibitem[Javanainen et~al.(1988)]{Java88}
J.~Javanainen, J.~H. Eberly, and Q.~Su, \emph{Phys.~Rev.~A} \textbf{38},
  3430--3446 (1988).

\bibitem[Casati et~al.(1988)]{Casa88}
G.~Casati, I.~Guarneri, and D.~Shepelyansky, \emph{IEEE} \textbf{24},
  1420--1444 (1988).

\bibitem[Blumel and Reinhardt(1997)]{ChaosAtomPhys}
R.~Blumel, and W.~P. Reinhardt, \emph{Chaos in Atomic Physics}, Cambridge U.
  Press, 1997.

\bibitem[Buchleitner et~al.(2006)]{Buch06}
A.~Buchleitner, {\it et al.}\ 
  \emph{Phys.~Rev.~Lett.} \textbf{96},
  164101 (2006).

\bibitem[Cvitanovi\'{c} et~al.(2008)]{chaosbook}
P.~Cvitanovi\'{c}, {\it et al.}\ 
  \emph{Chaos: Classical and Quantum}, Niels Bohr Institute, Copenhagen, 2008,
  {\tt {http://ChaosBook.org}{ChaosBook.org}}.

\end{thebibliography}


\end{document}